                        The Latex File

\documentstyle[12pt]{article}
\begin{document}
\newcommand{\myappendix}{\setcounter{equation}{0}\appendix}
\newcommand{\mysection}{\setcounter{equation}{0}\section}
\renewcommand{\theequation}{\thesection.\arabic{equation}}

\def\be{\begin{equation}}
\def\eq{\end{equation}}
\def\bea{\begin{eqnarray}}
\def\eea{\end{eqnarray}}

\topmargin=.17in			
\headheight=0in				
\headsep=0in			
\textheight=9in				
\footheight=3ex				
\footskip=4ex		
\textwidth=6in				
\hsize=6in				
\parindent=21pt				
\parskip=\medskipamount			
\lineskip=0pt				
\abovedisplayskip=3em plus.3em minus.5em 
\belowdisplayskip=3em plus.3em minus.5em	
\abovedisplayshortskip=.5em plus.2em minus.4em	
\belowdisplayshortskip=.5em plus.2em minus.4em	
\def\baselinestretch{1.2}	
\thicklines			    
\oddsidemargin=.25in \evensidemargin=.25in	
\marginparwidth=.85in				

\newcommand{\comb}{\mbox{$\displaystyle{g^2 \mu^{2 \epsilon} \over 2}$}}
\newcommand{\combo}{\mbox{$\displaystyle{g^4 \mu^{4 \epsilon} \over 3!}$}}
\newcommand{\meas}{\mbox{$\displaystyle{ \int {d^{D-1} k \over
(2\pi)^{D-1}}}$}}
\newcommand{\mead}{\mbox{$\displaystyle{  {d^{D-1} k \over (2\pi)^{D-1}}}$}}
\newcommand{\covk}{\mbox{$\displaystyle{\int  {d^{D} k \over (2\pi)^{D}}}$}}
\newcommand{\covq}{\mbox{$\displaystyle{\int  {d^{D} q \over (2\pi)^{D}}}$}}
\newcommand{\meaq}{\mbox{$\displaystyle{ \int {d^{D-1} q \over
(2\pi)^{D-1}}}$}}
\newcommand{\meak}{\mbox{$\displaystyle{ {d^{D-1} q \over (2\pi)^{D-1}}}$}}
\newcommand{\denom}{\mbox{$\displaystyle{1 \over 8 E_{k} E_{q} E_r}$}}
\pagestyle{plain}

\baselineskip=16pt

\hfill{ITP-SB-91-64}\\
\vskip 0.2cm
\begin{center}
{\Large\bf Resummation in a Hot Scalar Field Theory}\\
\vspace{0.5in}
Rajesh R. Parwani \\
\end{center}
\vspace{.10in}
\centerline{\it Institute for Theoretical Physics \footnote
{email:rparwani@max.physics.sunysb.edu\\
Adress after
Sept.1 1992: Service de Physique Theorique de Saclay, 91191 Gif-Sur-Yvette
Cedex,
France.}}
\centerline{\it State University of New York at Stony Brook}
\centerline{\it Stony Brook, NY \ \ 11794-3840}
\vspace{0.7in}
\centerline{\sc December 1991}
\centerline{ Revised}
\centerline{\sc March 1992}
\vspace{0.5in}
\centerline{\large\bf Abstract}

A resummed perturbative expansion is used to
obtain the self-energy
in the high-temperature \(g^2\phi^4\) field theory model
up to order $g^4$. From this the zero momentum pole
of the effective propagator is  evaluated  to determine
the induced thermal mass and damping rate
for the bosons in the plasma to order $g^3$.  The calculations are
 performed in the imaginary time formalism and
a simple diagrammatic analysis is used to identify the
relevant diagrams at each order. Results are compared with similar
real-time calculations found in the literature.

\newpage

\mysection{Introduction}
A well known [1] problem in high-temperature ($T$) field theory is the
breakdown of the conventional perturbative expansion at some order in
the coupling constant ($g$). This happens because in the regime
 $T \gg gT \gg m_{0}$, where $m_{0}$ represents
any intrinsic zero-temperature masses in the theory, the relevant cutoff for
infrared (IR) singularities in loop diagrams is the thermal mass
($\sim gT$) rather than $m_0$.
Higher loop diagrams then accumulate powers of $g$ in
the denominator which can compensate for the usual factors of $g$ in the
numerator coming from the Feynman rules.
Therefore, to compute consistently to a given order in $g$,
we have to take into account all the relevant higher loop graphs---these
usually form an infinite set.

A practical solution is to resum the perturbation
series by systematically including [2,3]  all lower order radiative
corrections that are significant (like the thermal mass)
in higher order calculations.
For gauge theories, the required resummation of the perturbative expansion
into an effective expansion was developed  recently
  by Braaten and Pisarski [3] to compute the gluon damping rate to leading
( $\sim g^2T$) order. Subsequently, the
 effective
expansion has been used to compute many other quantities [4].
In all these applications , only one-loop diagrams in the {\em effective}
 theory were considered.

To go beyond leading order, one must compute two-loop
(and higher) diagrams in the effective
expansion. Since this is a tedious exercise in gauge theories, I will
in this paper
deal with a toy model---the $g^2 \phi^4$ theory---in order to explore some of
the technical aspects of higher loop calculations within the resummation
program. As will be discussed in Sec.2, for this model only the self-energy
has to be resummed while the vertex can still be treated perturbatively as in
the bare theory [3]. A two-loop calculation in the same model with partial
resummation has been considered by Altherr [5]. More recently,
a modified perturbation expansion for the model was
proposed by Banerjee and Mallik [6] to enable the systematic
 calculation of the effective mass to
higher orders. In [6] a mass parameter was introduced in the beginning
 and later determined by consistency conditions.

The main difference between [5,6] and this paper is that here the
imaginary-time formulation (ITF) will be used to perform the calculations
whereas the real-time formulation (RTF) was employed in [5] and [6].
In the ITF
the diagrammatics is the same as at $T = 0$ and the
power-counting of IR divergences is extremely simple. These
conviniences of the ITF will be exploited to give a careful account of all
the diagrams that can contribute to a given order in $g$ towards
the self-energy. Also, instead of introducing a mass parameter as in [6],
the resummation
will be done in stages so as to make it easier to  identify
the relevant diagrams and ranges of momenta which
can contribute to a particular order in the coupling constant.
As an example of an explicit calculation, I will
determine the thermal mass and damping rate, for bosons at zero momentum,
up to order $g^3$. The results will be compared with those obtained in the
RTF. Of course, as the scalar model is quite popular, some of the
formulae and results obtained in this paper, especially in Sec.2,
 may be found in other publications [3, 5-8 ].

The plan for the rest of the paper is as follows : in Sec.2 I will set up the
notation and perform the first stage in the resummation of self-energy
diagrams. This includes
only one-loop one-particle irreducible (1PI) diagrams.
At this stage, the self-energy is momemtum independent
so the  induced thermal mass is easily obtained to order $g^2$.
Although  individual higher-loop
1PI self-energy  diagrams in the effective expansion seem to
contribute to this order, it is demonstrated that their {\em sum} does not.
Thus perturbative computability is maintained in the effective expansion.
The one-loop 4-point function
is also considered in order to explain why the vertex corrections can be
treated perturbatively.
In Sec.3, the effective
lagrangian of the previous section is used to perform the next stage of
the resummation, which includes both one and
two-loop diagrams. The thermal mass is obtained  up to order
  $g^3$. Again, the sum of higher-loop diagrams is shown to
 cancel at this order.
The imaginary part of the self-energy is also computed to determine
the damping rate to lowest order.
The conclusion and a summary is in Sec.4, while the appendix contains some
technical details.

\mysection{One-loop}
The starting point is the following  lagrangian for a hot scalar field
(i.e. the intrinsic mass has been set to zero)

\be
{\cal L}_{0} = \frac{1}{2}(\partial_{\mu}\phi)^2 + \frac{g^2
\mu^{2\epsilon}}{4!} \phi^4
\eq

The lagrangian has been written in $D$ dimensional Euclidean space, where
 $D =4-2\epsilon$ and
$\mu$ is the mass parameter of dimensional regularization. The
renormalization  counterterms, which have not been displayed,
will be determined in the minimal subtraction scheme [9].
In the imaginary time formulation of
finite-temperature field theory [10,11], the information about the temperature
($T$) is
contained in the  energies which are
now discrete; for bosons, $p^0 = 2\pi jT$ , where $j$ is an integer.
The only change from
the zero-temperature Feynman rules is  then in the replacement (much
of the notation is similar to [12] )

\be
\int d^{D}k / (2 \pi)^{D} \longrightarrow \mbox{Tr}_{k} \equiv T \sum_{j= -
\infty}^{+ \infty} \int d^{D-1}k / (2 \pi)^{D-1}
\eq

The sum over the discrete frequencies inside loops is most
efficiently performed using the ``Saclay''
method [12]. Real-time amplitudes are then obtained by analytically
continuing[10]  the external energies, $p^0 \rightarrow -i\omega$.
Let $\Delta(K)$ represent a bosonic propagator with mass $M$
and momentum $K^2 = (k^{0})^{2} + k^{2}$ ,
\be
\Delta(K)= \frac{1}{K^2 + M^2} .
\eq
For the lagrangian ${\cal L}_0$, the massless propagator will be
denoted as  $\Delta_{0}(K) = 1/K^2$.
In the Saclay method, the propagators inside loops are replaced by
their spectral representations

\bea
\Delta(K) &=&  \int_{0}^{1/T} d \tau \ e^{ik^{0}\tau}  \ \Delta(\tau,k)
\nonumber\\
\nonumber \\
\Delta(\tau,k) & = & (1/2E_{k})[(1+n_{k})e^{-E_{k}\tau} + n_{k} \ e^{E_{k}
\tau}]
\eea
Here, $E^{2}_{k}=k^{2}+M^{2}$ and $n_{k} =1/(\mbox{exp}(E_{k} /T) - 1)$ is the
Bose-Einstein distribution function. The expression (2.4) for the
noncovariant propagator is valid for $0\leq \tau \leq 1/T$ and
is defined to be periodic in $\tau$ with period $1/T$ outside that
range. By using
the spectral representation of the propagators, it is trivial to do the
 frequency sums
followed by the $\tau$ integrals, leaving only the integrals
over spatial momenta to be performed [12].

The main calculations in this paper will focus on obtaining consistently
the pole of the effective propagator, $1/(P^2 - \Pi(p^0 , \vec{p}))$,
where $P^\mu = (p^0,\vec{p})$ is the external 4-momentum and $\Pi$ is the
1PI self-energy.
For the theory described by (2.1), the diagram in Fig.1a can now be evaluated
as described above to determine the self-energy to lowest order

\bea
 \Pi_{0}(p^{\mu}) &=& - \comb\ \mbox{Tr}_{k} \ \Delta_{0}(K) \nonumber\\
\nonumber \\
& = & - \comb \left( \meas\ {1 \over 2k} \ + \  2\meas {n_{k} \over 2k}
 \right).
\eea

The first integral in (2.5) is the self-energy at $T=0$. It vanishes in
dimensional regularization [9], so there are no ultraviolet (UV)
divergences to this order.
The second integral in (2.5) represents the matter contribution and is UV
finite because of the Bose-Einstein (BE) factor. Putting $\epsilon=0$ then
gives the result

\be
\Pi_{0}(p^{\mu})  =  -{g^2 T^2 \over 24}.
\eq

In this paper, the induced thermal mass, $m$, is defined as the real part of
the
pole of the Minkowski  propagator at zero momentum
($\vec{p} = 0$). Since the self-energy to this order is independent
of momentum, one gets
\be
m^2 \equiv  - \Pi_{0} = {g^2 T^2 \over 24}.
\eq

To systematically include the effects of this thermal mass, the term
$\frac{1}{2} m^2 \phi^2 $ is added and subtracted [2,3] from (2.1) to define
a new effective lagrangian

\be
{\cal L}_{2} = ( {\cal L}_{0} + \frac{1}{2} m^2 \phi^2 ) - \frac{1}{2} m^2
\phi^2.
\eq

The subscript `2' on ${\cal L}_{2}$ is used to remind us that the new
lagrangian now describes a theory with tree level mass $m$ with $(m/T)^2 \sim
g^2$.
In (2.8), the quantity in brackets defines a lagrangian with
free propagator $\Delta_{2}(K) = 1/ (K^2 +m^2)$. The subtracted term is treated
as
a new 2-point interaction (Fig.1c) of order $g^2$. The shifting of terms in
${\cal L}_{0}$ to form ${\cal L}_{2}$ corresponds to a resummation of the
perturbative expansion.

The next step [3] is to use the effective lagrangian (2.8) to recalculate the
self-energy. In addition to Fig.1a,  a contribution from the
new vertex (Fig.1c) must also be included,

\bea
\Pi_{3}(p^{\mu}) &=& m^2 - \comb \mbox{Tr}_{k} \ \Delta_{2}(K) \nonumber \\
\nonumber \\
 & = & m^2 - {g^2 \over 2} {m^2 \over (4\pi)^2} \left({4\pi \mu^{2}
\over m^2} \right)^
{\epsilon}\ \Gamma (-1 + \epsilon) - \comb \meas {2n_{k} \over 2E_{k}}
\eea
where now $E_{k}^{2} = k^2 + m^2$.
The second term in (2.9) is divergent as $\epsilon \rightarrow 0$. This UV
divergence is similar to that in $T=0$ field theory. The only difference
is that as a consequence of the resummation the thermal mass has
been introduced
into the perturbative calculations. This makes the above  divergence
 temperature dependent [6], albeit in a trivial way---the structure of the
divergence is the same as at $T=0$ , with the intrinsic mass $m_{0}$ replaced
by the thermal mass $m$. Therefore the structure of the mass counterterm will
still be the same as for $T=0$, ensuring that the theory is
renormalisable even though the counterterms are temperature dependent.
(See however the discussion following (2.17)).
 Expanding the divergent term near $ \epsilon = 0$ gives

\be
{g^2 \over 2}{m^2 \over (4 \pi)^2}{1 \over \epsilon} + \mbox{finite terms }
\vartheta (g^2 m^2 \ln{\mu^2 \over m^2})
\eq
where, since $m^2 \sim (gT)^2$, the finite term is $\vartheta (g^4 \ln g)$.
The mass counterterm vertex (Fig. 1b) is thereby fixed to be $-{g^2 m^2}/{32
 \pi^2 \epsilon}$ at lowest order. The one-loop renormalised self-energy in the
theory defined by
${\cal L}_{2}$ is then (See A.4)

\bea
\Pi_{3}^{ren}& =& m^2 - {g^2 \over 4 \pi^2} \int_{0}^{\infty} dk {k^2 n_{k}
\over E_k} \nonumber \\
\nonumber \\
& =& 3m^3/{\pi T} + \vartheta(g^4 \ln g).
\eea

The corrected thermal mass, $M_{3}$, is given by,

\be
M_{3}^{2} = m^2 - \Pi_{3}^{ren} = m^2(1-3m/{\pi T})\ + \ \vartheta(g^4 \ln g).
\eq

Calculating Fig.1a using the propagator $\Delta_{2}(K)$
is equivalent to summing the infinite
set of ``daisy'' [13] diagrams of Fig.2  evaluated with the massless propagator
$\Delta_{0}(K)$. This interpretation follows once $\Delta_{2}(K) =1/(K^2 +
m^2)$
is expanded in a Taylor series in $m^2$ about $m^2=0$. Each
of the diagrams of Fig.2 (for $N \geq 1$)
is infrared divergent in the theory ${\cal L}_{0}$ but their sum is,
 as we have seen, infrared finite.
Thus summing an infinite set of IR divergent diagrams has given an
IR finite correction of order $g^3$ to the mass. The  nonanalytic (in $g^2$)
behaviour of this correction is a sign of its nonperturbative nature
(infinite resummation) when
viewed in terms of the original lagrangian (2.1) [5,11].

Are there any other diagrams in the effective theory (2.1) which can
contribute terms of order $g^3$ to the self-energy ? The answer, at first
sight, is yes.
Even in the {\em effective} theory, there
are {\em infinitely} many diagrams, other than those in Fig.1, which can
contribute at
order $g^3$ but, fortunately for the consistency of the resummation,
their {\em sum} is
of order $g^4$ or higher! Let me term such  diagrams
``irrelevant'' since eventually their finite contributions
to the present order in $g$ cancels , though they might be relevant
for the UV renormalisation of the theory.

Examples of irrelevant
diagrams at order $g^3$ are given in Fig.3. Each of the diagrams there is
$\vartheta (g^2)^2({1 \over g}) \sim g^3$. The factor $(g^2)^2$ comes from the
vertices while the $1/g$ comes from the bottom loops as their IR singularity
is cutoff by the thermal mass. The simplest way to deduce the factor of $1/g$
is to note that the IR behaviour of bosonic propagators in loops
is dominated by the $j=0$ term in the frequency sum (2.2). That is , to get
the leading IR behaviour of a diagram,  set all the internal energies to zero
and then take the $m  \rightarrow 0$ limit. Consider for
example Fig.3a. Its IR behaviour is

\bea
g^4 \mbox{Tr}_{k}  \mbox{Tr}_{q} \left(\Delta_{2}(K) \left[\Delta_{2}(Q)
\right]^2 \right) &
\sim & g^4
\int {d^3 k \over k^2 + m^2}\int {d^3 q \over (q^2 + m^2)^2} \nonumber \\
\nonumber \\
& \rightarrow &\vartheta (g^4)(1)(1/g) = \vartheta (g^3).
\eea
However, the sum of graphs in Fig.3 ,with the proper combinatorial factors, is
(using eqn.(2.11))
\bea
\comb \ \mbox{Tr}_{q} \left[\Delta_{2}(Q) \right]^2 \left(\comb \
\mbox{Tr}_{k} \Delta_{2}(K)\ -\ m^2 \right) &=& - \comb \ (\Pi_{3})
\mbox{ Tr}_{q} \left[\Delta_{2}(Q)\right]^2 \nonumber \\
\nonumber \\
& \sim & \vartheta (g^2)(g^3)(1/g) = \vartheta(g^4).
\eea
The UV divergent parts of course cancel only when all the relevant two-loop
 and counterterm diagrams are summed ( Sec.3). Similarly, although
each of
the daisy-like diagrams shown in Fig.4 is $\vartheta (g^2)^3 (1/g^3) \sim g^3$,
their
sum is easily shown to be
\be
- \comb  \ \mbox{Tr}_{q} \left[\Delta_{2}(Q)\right]^3 \left(\comb \
\mbox{Tr}_{k} \Delta_{2}(K)\ -\ m^2 \right)^2 \ \sim \vartheta
(g^2)(1/g^3)(g^3)^2 =
\vartheta(g^5).
\eq

In general each of the daisy-like diagrams in Fig.5 with a fixed number $N\geq
 2$ of ``bubbles'' (Fig.1a) $+$ ``blobs''(Fig.1c) is $\vartheta (g^3)$ but
their sum is

\bea
& &-{g^2 \mu^{2\epsilon} } \ \mbox{Tr}_{q} \left[\Delta_{2}(Q)\right]^{N+1} \
\sum_{p=0}^{N} {\left[ -g^2 \mu^{2 \epsilon} \mbox{Tr}_{k} \Delta_{2}(K)
\right]^{p} \over2^p p!} { (m^2)^{N-p} \over (N-p)!} \nonumber \\
\nonumber \\
 & &  = -{g^2 \mu^{2\epsilon} \over N!} \ \mbox{Tr}_{q} \left[\Delta_{2}(Q)
\right]^{N+1}
\left(m^2 - \comb \ \mbox{Tr}_{k} \Delta_{2}(K)\right)^N \nonumber \\
\nonumber \\
 & & \sim \vartheta  (g^2)(1/g^{2N-1})(g^{3N}) = \vartheta(g^{N+3}).
\eea
Thus the set of all daisy-like  diagrams with $N \geq 2$ is completely
irrelevant
for the calculations in this paper which will be performed up to
order $g^4$. It is clear from the above analysis that the presence of the
2-point interaction (Fig.1c) is essential. Recall that it was
introduced (2.8) to keep us in the same fundamental theory while performing
the resummation. We see now how, in the cancellation of contributions from
the infinite set of daisy diagrams, it prevents an overcounting of diagrams.

The $N=1$ daisies of Fig.3 which seem to be relevant at order $g^4$ will
be discussed further in the next section. It is left as an exercise for the
interested reader to verify, using the simple power counting rules for IR
singularities illustrated above, that any other 1PI self-energy diagram
is individually of order $g^4$ or higher.


To summarise, the thermal mass-squared including all subleading corrections of
order $g^3$ is completely given by (2.12).

So far, all the results have been written in terms of the renormalised
coupling $g$. The `physical' coupling is determined by evaluating the diagrams
of Fig.6 on shell , which corresponds to soft ($ \sim gT$) external momenta.
Using the by now familiar power counting, it is seen that each of the diagrams
is $\vartheta (g^3)$ and hence the radiative correction to the basic 4-point
vertex is down by a factor of $g$ [7]. Therefore,
vertex corrections can be treated perturbatively instead of resumming the
corrections to form effective 4-point vertices. Contrast this with the
thermal mass , $m$, which is of the same order as the bare inverse
propagator  $\Delta_{0}^{-1}(K)$ for soft momenta and therefore has
to be resummed. In the language of [3], for the scalar theory, the only
`` hard thermal loops'' are in the self-energy. In this paper, all the results
will be left in terms of the renormalised coupling $g$.

To obtain the complete effective lagrangian to order $g^3$ the
vertex renormalisation counterterm is needed. This is determined as usual
 by calculating
Fig.(6a) ( plus the usual crossed diagrams ) at T=0. Including first all the
counterterms in (2.8) gives

\bea
{\cal L}_{2} \rightarrow {\cal L}_{2}^{\prime}  &=&  ({\cal L}_{0} \ + \
m^2 \phi^2 /2) \  + \ {\phi^2 \over 2!}\left({g^2 m^2
\over (4 \pi)^2} {1 \over 2\epsilon}\right) \ +
\ {g^2 \mu^{2\epsilon}
\over 4!} \phi^4 \left({3g^2 \over (4\pi)^2}{ 1 \over
2\epsilon}\right) \nonumber \\
\\
& & \; - \left( \ m^2 \phi^2 /2 \ + \ {\phi^2 \over 2!}{g^2 m^2
\over (4 \pi)^2} {1 \over 2\epsilon}\right). \nonumber
\eea

Included in ${\cal L}_{2}^{\prime}$ is the counterterm (Fig.7a) for loop
corrections to the 2-point interaction. Note that the net lagrangian
does not contain temperature dependent counterterms though pieces of it do
because of the resummation [16].

Then, as before, the effects of the thermal mass
to order $g^3$ are included by shifting the mass term in (2.17),

\bea
{\cal L}_{3}  &=& ( {\cal L}_{0} \ + \ M_{3}^{2} \phi^2 /2) \
 + \ {\phi^{2} \over 2!} \left({g^2 M_{3}^{2} \over (4 \pi)^2} {1 \over
2\epsilon}\right) \ + \ {g^2 \mu^{2\epsilon} \over 4!} \phi^4
\left({3g^2 \over (4\pi)^2}{ 1\over 2\epsilon}\right) \nonumber \\
\\
& & \; - \left( \ M_{3}^{2} \phi^2 /2 \ + \ {\phi^2 \over 2!}{g^2 M_{3}^{2}
\over (4 \pi)^2} {1 \over 2\epsilon}\right). \nonumber
\eea

Notice that for consistency, the mass in the UV counterterms has also been
 shifted to $M_{3}$ in order to cancel the divergences in loop calculations.
The lagrangian (2.18) will be used in the next section to obtain
the self-energy up to order $g^4$. Strictly speaking, the further resummation
to obtain ${\cal L}_{3}$ is unnecessary as the $\vartheta(g^3)$ correction
to $m^2$ is a perturbative correction, just like the $\vartheta(g^3)$
correction to the 4-point vertex. However, no harm is done by this additional
resummation, all that happens is a redistribution among the diagrams
of the next correction at order $g^4$, as we will soon see.

\mysection{Two-loop}

The basic diagrams that must be considered to evaluate the self-energy
to order $g^4$ using ${\cal L}_{3}$ are in Figs.1,3,7 and 8. Before
delving into the calculations, let us make some observations.
Just as for $T=0$, the sum of graphs in Fig.1  must be UV finite. Also, as
for $T=0$,
Figs.7b and 7c are the mass and vertex counterterm diagrams that are required
to cancel the subdivergences arising from the two-loop diagrams Fig.3a and
Fig.8 . A `new' [6] feature of the resummation
is  the counterterm of Fig.7a which is needed to cancel UV divergences
generated by loop corrections (Fig.3b) to the 2-point vertex (Fig.1c).

The sum of diagrams in Fig.1, evaluated from the lagrangian ${\cal L}_{3}$
is,

\bea
\Pi_{4}^{(1)}(p^{\mu}) &=& M_{3}^2 \ - \ \comb \mbox{Tr}_{k} \ \Delta_{3}(K)
\ - \ {g^2 \over 2}{M_{3}^2 \over (4 \pi)^2}{1 \over \epsilon} \nonumber \\
\nonumber \\
 &=& M_{3}^2 - m^2 \left[1-{3m \over \pi T} -{3 \over 4\pi^2}\left({m \over T}
\right)^2 \ln \left({m \over T}\right)^2 - {3 \over \pi}\left({m \over T}
\right)^2 \left(C_{1} -{3 \over 2\pi}\right)\right] \nonumber \\
\nonumber \\
& & \; + {1 \over 2}  \left({gm \over 4\pi}\right)^2 \left[\ln \left({4\pi
\mu^2 \over m^2} \right) + (1 -\gamma_{E})\right] \ + \ \vartheta(g^5 \ln g),
\eea
where $\gamma_E$ is the Euler constant and $C_{1}={1 \over 4}(2\gamma_E -2\ln
4\pi
 - \ 1)$. Eqns.(2.12) and (A.4) were used to get the final form (3.1).

Fig.3 contributes

\be
\Pi_{4}^{(3)}(p^\mu) = \comb \ \mbox{Tr}_{q}
\left[\Delta_{3}(Q)\right]^2 \left(\comb \ \mbox{Tr}_{k} \Delta_{3}(K)\
-\ M^2_{3}\right).
\eq
Since $ (g^2/2)\mbox{Tr}_{k} \Delta_{3}(K) = M_{3}^2 + \vartheta(g^4 \ln g)$,
therefore the finite part of (3.2) is
$\vartheta (g^5 \ln g)$. In the last section, working with ${\cal L}_{2}$,
it was shown that the same diagrams  sum to $\vartheta (g^4)$. The sum has now
been
pushed to higher order because of the further resummation  performed to
obtain ${\cal L }_{3}$. This is an example of the
`redistribution' mentioned at the end of the last section ---the `lost'
contribution from Fig.3 has been picked up by Fig.1a: this is indicated in
(3.1) by the presence of the factor $(C_{1} -3/2 \pi)$ rather than $C_{1}$,
the latter factor being the contribution if ${\cal L}_{2}$ were used in
calculating Fig.1.
The diagrams of Fig.3 are thus only needed to
complete the UV renormalisation of the theory which, as usual, is performed
loop-wise.

The only other graph that is relevant for  discussion is given in Fig.8;
it will be considered later.
The daisy-like diagrams (with $N \geq 2$) were already shown in the last
section
 to be irrelevant at $\vartheta( g^4)$. However now a new infinite set of
 graphs must be analysed, the simplest of which are shown in Fig.9. Each of
the diagrams in Fig.9 is of order $g^4$ by power counting but their sum is
clearly $\vartheta (g^6 \ln g)$. Extending diagram (9a) by adding
a bubble (or blob)
to its top gives a graph of order $g^5$. So one only needs to consider adding
$N$ number of bubbles $+$ blobs to the middle loop and $M$ bubbles $+$ blobs to
the
bottom loop of Fig.9a to create the general `` cactus'' diagram of
order $g^4$ shown
in Fig.10. It is sufficient to show that the sum
of all such cactus diagrams is of order $g^5$ or higher :
  first consider Fig.10 with the bottom loop and its $M$ attachments
fixed in a particular configuration. Then any subdiagram (with fixed $N \geq
0$) above the bottom loop is precisely a daisy diagram and these have been
shown to sum to $\vartheta (g^5)$ at most. For any $M \geq 0$ the bottom loop
in Fig.10 contributes a factor $g^2(1/g) =g$. Hence the sum of all possible
cactus diagrams is at most of order $g^6$. This completes the proof.

Adding bubbles or blobs to Fig.8 creates diagrams like  those shown
in Fig.11. These
sum to order $g^6 \ln g$. All other diagrams are individually of order
$g^5 \ln g$ or higher.

Having accounted for all the relevant diagrams, let us return to
some explicit results. The counterterms in Fig.7 contribute

\bea
\Pi_{4}^{(7)} &=& \left({gM_{3} \over 4\pi}\right)^2 {1 \over 2\epsilon} \ + \
 \comb \left[\left( {gM_{3} \over 4\pi} \right)^2 {1 \over 2\epsilon}\right]
 \mbox{Tr}_{k} \left[\Delta_3(K)\right]^2 \nonumber \\
\nonumber \\
& & - \comb \left[{3g^2 \over (4\pi)^2} {1 \over 2\epsilon}\right]
\mbox{Tr}_{k}\Delta_{3}(K).
\eea
Summing (3.2) and (3.3) gives

\bea
\Pi_{4}^{(3,7)} &=& {-g^4 \over 2\epsilon}{ I_{\beta}^{\epsilon}(M_{3})
 \over (4\pi)^2} \nonumber \\
\nonumber \\
& & - {g^4 \over 2\epsilon}{M^2_{3} \over (4\pi)^4} \left[ (\gamma_{E}-1)
- \ln \left({4\pi \mu^2 \over M_{3}^2}\right) \right] \nonumber \\
\nonumber \\
& &+ {3g^4 \over 4 \epsilon^2}{M_{3}^2 \over (4\pi)^4} + \ \mbox{finite terms}
\ \vartheta(g^5 \ln g),
\eea
where $ \displaystyle{ I_{\beta}^{\epsilon}(M_{3}) \equiv \mu^{2 \epsilon}
\meas {n_{k} \over E_{k}}}. $

The first line in (3.4) is a nonrenormalisable temperature dependent infinity
generated by diagrams (3a) and (7c). It will cancel [8] when the
 two-loop overlapping diagram of Fig.8 is added. This last diagram is the only
relevant diagram which depends on the external momentum $(p^0 ,\vec{p})$,

\bea
\Pi_{4}^{(8)}(p^0 , \vec{p})& =& {g^4 \mu^{4\epsilon}  \over 3!} \mbox{Tr}_{k}
\mbox{Tr}_{q} \left[\Delta_{3}(K) \Delta_{3}(Q) \Delta_{3}(P-K-Q)\right]
 \nonumber \\
\nonumber \\
&=& G_{0}(p^0,\vec{p}) + G_{1}(p^0, \vec{p}) + G_{2}(p^0, \vec{p})
\eea
where
\bea
G_{0}(p^0, \vec{p})& =& \int d[k,q] \ S(E_{k}, E_{q}, E_{r}) \\
\nonumber \\
G_{1}(p^0, \vec{p})& = & 3 \int d[k,q] \ n_{k} \left[ S(E_{k}, E_{q}, E_{r})
+ S(-E_{k}, E_{q}, E_{r}) \right]
\eea
\be
G_{2}(p^0, \vec{p}) = 3 \int d[k,q] \ n_{k} n_{q} \left[
S(E_{k}, E_{q}, E_{r}) + S(-E_{k}, E_{q}, E_{r}) +
S(E_{k},-E_{q}, E_{r}) - S(E_{k}, E_{q},-E_{r}) \right]
\eq
with the definitions \\
$$ S(E_{k}, E_{q}, E_{r}) = \left({1 \over ip^0 + E_k + E_q +E_r} +
 {1 \over -ip^0 + E_k + E_q +E_r} \right)$$
\begin{eqnarray*}
d[k,q] &=& \combo \mead \meak \denom\\
\\
 r & =& | \vec{k} + \vec{q} - \vec{p} | \\
\\
E_{l}^{2} &=& l^2 + M^{2}_{3} \ , \ l = k,q,r \\
\end{eqnarray*}
and $ n_{l}$ the usual BE factor.
The real-time retarded self-energy follows by making the analytic
continuation
 $p^0 \rightarrow -i\omega + \xi \mbox{ with } \xi = 0^{+}$ [10].
Then the prescription

\be
{1 \over A \pm i\xi} = \mbox{P}\left({1 \over A}\right) \mp i \pi \delta(A) \\
\eq
gives the real and imaginary parts of the diagram [14]. Consider first the
real part.
Since $G_0$ does not contain any Bose-Einstein factors, it must
be the  expression for diagram (8) obtained using $T=0$
Feynman rules and with the energy integrals done. In covariant (i.e. with
$P^2 = -\omega^2 + (\vec{p})^2$ ) notation one gets [15]

\bea
\mbox{Re } G_{0}(P^2) &=&\mbox{Re } \combo \covk \covq {1 \over K^2 + M_{3}^2}
{1 \over Q^2 + M_{3}^2} {1 \over (K+Q-P)^2 + M_{3}^2} \nonumber \\
\nonumber \\
&=& {-g^4 \over 4}{M_{3}^{2} \over (4 \pi)^4} \left[ {1 \over
\epsilon^2} + {3-2\gamma_{E} \over \epsilon} + { 2 \over \epsilon}\ln \left({4
\pi
 \mu^2\over M_{3}^{2}}\right) \right] - {g^4 \over 4}{P^{2} \over
(4 \pi)^4}{1 \over 6 \epsilon} \\
\nonumber \\
& & \  + \ \mbox{finite terms } \vartheta (g^4 M_{3}^2)({ P^2 / M_{3}^{2}})^2
\nonumber
\eea
For soft external momenta ($P^2 \sim m^2$),
the region of interest, the finite terms are of order $g^6$ and so do
not contribute to the self-energy at order $g^4$.

$G_{1}$ represents the mixing of the  $T=0$ piece from one
loop with the $T \neq 0$ piece from the second loop. This is clear from
the expression (3.7) which contains only one  BE factor, making one of
the loop integrals UV finite while the  other loop integral  has a UV
divergence. Specialising to the case $\vec{p}=0$ in order to do the angular
integrals, gives (See Appendix)

\bea
\mbox{Re } G_{1}(-i\omega ,0) &=& F_{0} +F_{1} +F_{2}(\omega^2) \\
\nonumber \\
\mbox{where } F_{0} &=& {g^4  \over 2}{I_{\beta}^{\epsilon}(M_{3}) \over
(4 \pi)^2}{1 \over \epsilon} \\
\nonumber \\
F_{1} &=& {g^4 \over 2}{I_{\beta}^{\epsilon=0}(M_{3}) \over (4 \pi)^2}
\left( \ln {4\pi \mu^2 \over M_{3}^{2}} + 2 - \gamma_{E} \right) \nonumber \\
\nonumber \\
&= & \left({gm \over 4\pi}\right)^2 \left( \ln {4\pi \mu^2 \over m^{2}} + 2 -
\gamma_{E} \right) + \vartheta (g^5 \ln g) \\
\nonumber \\
\mbox{and }
F_{2}(\omega^2) &=& {g^4 \over 8 (2 \pi)^4} \int_{0}^{\infty} dk {k n_{k}
\over E_{k}} \int_{0}^{\infty} {dq \over E_{q}} \left(q \ln \left| {X_{+}
\over X_{-}} \right| -4k \right)\\
\nonumber \\
\mbox{with } X_{\pm} &=& \left[ \omega^2 - (E_{k} + E_{q} + E_{k \pm q})^2
\right] \left[ \omega^2 - (E_{q} - E_{k} + E_{k \pm q})^2 \right].  \nonumber
\eea
The temperature dependent infinity $F_{0}$ is actually independent of
the external momenta $p^{\mu}$ and cancels precisely against a  similar
term found earlier in eqn.(3.4).

Finally, $G_{2}$ contains a BE factor for each loop and so is UV finite. It
however has a logarithmic IR divergence as $m ,\omega \rightarrow 0$. One
obtains

\be
H(\omega^2) \equiv \mbox{Re} G_{2}(-i \omega,0) = {g^4 \over 8(2 \pi)^4}
 \int_{0}^{\infty} dk {k n_{k}\over E_{k}} \int_{0}^{\infty} dq {q n_{q} \over
E_{q}} \ \ln \left| {Y_{+} \over Y_{-}} \right|
\eq
where
\bea
Y_{\pm}& =&  \left[ \omega^2 - (E_{k} + E_{q} + E_{k \pm q})^2
\right] \left[ \omega^2 - (E_{q} - E_{k} + E_{k \pm q})^2 \right] \nonumber \\
& & \mbox{ } \  \times \left[ \omega^2 - (E_{k} - E_{q} + E_{k \pm q})^2
\right] \left[ \omega^2 - (E_{k} + E_{q} - E_{k \pm q})^2 \right].
\eea

The sum of all UV divergent terms from eqns.(3.4, 3.10, 3.12) gives

\be
\left({g^2 \over 16 \pi^2} \right)^2  M_{3}^{2} \left( {1 \over 2 \epsilon^2}
- {1 \over 4 \epsilon} \right) - \left({g^2 \over 16 \pi^2} \right)^2 {P^2
 \over 24 \epsilon}.
\eq
These are cancelled by the two-loop wave-function and mass renormalisation
counterterms

\be
 - {1 \over 2}( \partial_{\mu} \phi)^2  \left[ \left({g^2 \over 16
\pi^2}\right)^2 {1 \over 24 \epsilon} \right] + {M_{3}^2 \over 2}\left[
\left({g^2 \over 16 \pi^2}\right)^2 \left({1 \over 2 \epsilon^2} -{1 \over
4\epsilon }\right) \right].
\eq
The temperature dependent UV mass counterterm in (3.18) is $\vartheta (g^6)$.
As before, it will precisely compensate [16]  the $\vartheta (g^6)$
temperature dependent UV counterterm  for the 2-point interaction (the
diagrams that require the latter counterterm are formed by adding a blob
to Figs. 3a , 7b, 7c and 8).

The real part of the renormalised self-energy for ${\cal L}_{3}$ is therefore
(3.1,13-15)
\be
 R(\omega^2) \equiv \mbox{Re } \Pi_{4}^{ren}(-i \omega ,0) = \Pi_{4}^{(1)} +
F_{1} + F_{2}(\omega^2) + H(\omega^2) + \mbox{terms of order }(g^5 \ln g).
\eq
This expression contains all corrections  at order $g^4$. It also contains
some effects at order $g^5$ and higher in the energy dependent terms $F_{2}$
and $H$. For general $\omega$, the expressions $F_2(\omega)$ and $H(\omega)$
are too complicated to evaluate in closed form. However since only
contributions to $\vartheta (g^4)$ are required, something can be said. Note
that because of the explicit factor of $g^4$ , it is only necessary to
identify the IR behaviour of the integrals in  eqns.(3.14-15) to obtain
information
about the leading contibution. Clearly, the $\omega$ dependence of
any IR  behaviour in $F_{2}$ or $H$ can only be possible
for $\omega $ soft ($\sim m$). Consider first $F_{2}(\omega^2)$. It is easy
to see that the logarithmic IR singularity as $m,\omega \rightarrow 0$ is due
to the
second factor in the term $X_{-}$ in the region $k \geq q$. So one is led to
investigate the piece

\be
 \int_{0}^{\infty} {xdx \over E_{x}} n_{x} \int_{0}^{x} {ydy \over E_{y}}
\ln \left[ (E_{y} + E_{x-y} - E_{x})^2 - \sigma^2 \right],
\eq

\noindent where I have factored out $T^2$ and defined the set of dimensionless
variables $ \{ x,y,a,\sigma \}$ by scaling the quantities $\{k,q,m,\omega\}$
respectively by $1/T$. Now, for $a \rightarrow 0$ the estimate \\
$(E_{y} + E_{x-y} -E_{x})^2 \sim \vartheta(a^4)(1/y + 1/(x-y) - 1/x)^2 $ holds.
 From this it can be
deduced that for $ \sigma \sim a^n \sim g^n$ the `leading log' contribution
from (3.20) goes like $2 \ln a^2 $ for $n > 2$ and like $n \ln a^2 $ for
$n < 2$. For $H(\omega^2)$, the logarithmic IR singularity is caused
by the extra BE factor
while the magnitude of its contribution is controlled by the
$\ln (Y_{+}/Y_{-})$ term. Writing

\be
{Y_{\pm} \over T^8} = -(8xy)^2(x \pm y )^2 \sigma^2 + (a^2-\sigma^2)^2 Z_{\pm}
\eq

\begin{eqnarray*}
\mbox{where }  Z_{\pm}& =& 16\left[(x \pm y)^2 (x^2 + y^2) + (xy)^2 + a^2(x^2 +
y^2 \pm xy)\right] \\
& &  +8(a^2-\sigma^2)(x^2 + y^2 +a^2 \pm xy) \\
& &  + (a^2 - \sigma^2)^2,
\end{eqnarray*}
shows that the first term in (3.21) dominates for $\sigma \sim a$,
while the transition to more complicated behaviour is again at $\sigma
\sim ga \sim g^2$.

To get the full $g^4$ dependence from $F_{2}$ and $H$,
the constant under the
leading $g^4 \ln g$ contribution is also needed. This can be done for
specific values of $\omega$ when it is possible to isolate clearly
the $\vartheta(g^4)$ pieces from
the partial higher order effects. A calculation, which is sketched in
the appendix, gives (with corrections at $\vartheta (g^5)$ omitted)
\bea
F_{2}(0) & = & \lambda {\pi \over \sqrt{3}} \nonumber \\
H(0) & = &  \lambda \left[\ln \left({m \over T}\right)^2  + 3.48871... \right]
\nonumber \\
\\
F_{2}(M_{3}^2) & = & \lambda \left[- {1 \over 2} \ln \left({m \over T}\right)^2
 + 0.54597... \right] \nonumber \\
H(M_{3}^2) & = & \lambda \left[ {3 \over 2} \ln \left({m \over T}\right)^2
 + 4.52097... \right] , \nonumber
\eea

from which follows
\bea
 F_{2}(0) + H(0)& =& \lambda \left[ \ln \left( {m \over T}\right)^2 + 5.3025...
 \right]
\nonumber \\
\\
 F_{2}(M_{3}^2) + H(M_{3}^2)& =& \lambda \left[ \ln \left( {m \over T}\right)^2
+ 5.0669... \right], \nonumber
\eea
where $\lambda = - (gm / 4 \pi )^2$.
Surprisingly, from (3.23 ) it appears that the coefficient in front of the
total $\vartheta( g^4 \ln g)$
 contribution to the real self-energy (3.19) from
the energy-dependent part is the same on-shell as for zero external
4-momentum. Let me now proceed to the determination of the pole of the
effective propagator. The complex pole, $\Omega$, at zero-momentum ($\vec{p}
=0$)
is the zero of the equation

\be
-\Omega^2 + M_{3}^{2} - \Pi_{4}^{ren}(-i\omega,0) = 0.
\eq

Since Im $\Pi_{4}(-i \Omega , 0)$ is $\vartheta (g^4)$ (see later),
then by writing $\Omega = \omega - i \gamma$, the  real part of the pole
up to $\vartheta( g^4)$ is determined by

\be
-\omega^2 + M_{3}^{2} - R(\omega^2) = 0.
\eq

The above equation may be solved by iteration (See Appendix) to give the
thermal mass $M_{4}$ up to order $g^4$,

\be
M_{4}^{2} = M_{3}^{2} - R(M_{3}^{2}),
\eq
where the right hand side must be expanded up to order $g^4$. Using the values
for $F_{2}(M_{3}^{2})$ and $H(M_{3}^2)$ given in (3.23), together with eqns.
 (3.1), (3.13) and (3.19), in (3.26) above gives the final answer

\be
M_{4}^{2} = m^2 \left(1-{3m \over \pi T}\right) + \left({gm \over 4 \pi}
\right)^2 \left[{3 \over 2} \ln {T^2 \over 4\pi \mu^2} +2 \ln\left({m \over T}
\right)^2\right] +
\alpha \left({gm \over 4\pi}\right)^2
\eq
where $\alpha = 14.1416...\dots$ and $m^2$ is defined by eqn.(2.7). Taking the
square root of the above expression gives the complete thermal mass up to
order $g^3$.

The imaginary part [14] of the self-energy to order $g^4$ is due only to the
two-loop diagram of Fig.8. From (3.5-9), it is relatively simple to
obtain the imaginary part at zero-momentum and on-shell ,

\bea
\mbox{Im } \Pi_{4}^{(8)}(-iM_{3} ,0) &=& {g^4 \over 16}\left({1 \over 2\pi}
\right)^3
\int_{0}^{\infty} dk {k \eta_{k} \over E_{k}} \int_{0}^{k} {q dq \over E_{q}}
\nonumber \\
&=& {g^2 m^2 \over 32 \pi} + \vartheta (g^5).
\eea
The result, as expected on general grounds [12], is positive. The imaginary
part could also have been obtained directly, without using the
prescription (3.9), by keeping the full logarithm in (3.14-15) instead of only
its principal value. Finally, the damping rate  follows from (3.24),

\bea
\gamma &=& {\mbox{Im } \Pi_{4}^{(8)}(-iM_{3} , 0) \over 2M_{3}} \nonumber \\
 &=& {g^2 m \over 64 \pi} + \vartheta(g^4).
\eea

\mysection{Conclusion}

To summarise, the effective expansion created by a resummation in the
original theory was used to obtain the zero momentum pole of the propagator
in the energy plane
consistently to order $g^3$ .
 Working order by order in the effective expansion,
the relevant diagrams were identified and perturbative computability was shown
to hold by explicitly verifying the cancellation of contributions from an
infinite class of diagrams.

The first resummation of self-energy diagrams to get the effective lagrangian
${\cal L}_{2}$ was essential because the thermal mass at lowest order, $m$,
is as
large as the inverse massless propagator at soft momenta. That is, the
thermal mass could not be treated as a perturbation. As pointed out in the
text, the second resummation to form ${\cal L}_{3}$ was not really necessary
since the order $g^3$ correction to $m^2$ is a perturbative effect. The
consequence of the second resummation was simply to change the individual
contributions from some of the diagrams at $\vartheta (g^4)$. In particular,
whereas the diagrams in Fig.3 would have been relevant if we had continued
using ${\cal L}_{2}$, they became irrelevant when ${\cal L}_{3}$
 was used---this `lost' contribution was compensated by new
subleading contributions from the
diagrams in Fig.1. In short, though the reader could have been spared any
mention of
${\cal L}_{3}$, nevertheless the author feels that some insight into the
effects of resummation was gained by the exercise.

Let me now make some comparisons with results in the literature. In [6],
a mass parameter was introduced in the beginning and the effective mass was
defined by requiring that the corrections to the free inverse propagator
vanish at zero external 4-momenta. Translating that into the language of this
paper, simply amounts to using $R(0)$ on the right-hand side of (3.26)
rather than $R(M_{3}^2)$. In general, by definition, this  does not give the
pole in the propagator. From (3.23) we see that the difference
between the two definitions
shows up in the constant under the logarithm at order $g^4 \ln g$
(in the appendix I explain the difference).
I  do not know  however whether it is a coincidence that the coefficients of
the `logs'
in (3.23) are the same. The imaginary part of the self-energy of course
 vanishes for $\omega =0$, as
is apparent from (3.5-9) or from more general arguments [12], but is given
on-shell by (3.28).

Clearly, any quantity calculated in either the RTF
or ITF must give the same result
even though some of the intermediate expressions may look different because
of the differences in approach. For completeness, I have checked (using
 the real-time
expressions found in [5] and [6]) up to order $g^4$ that the two formalisms
give identical answers for  the pole of the
propagator; and also when used  to calculate the effective mass as defined in
[6].

For gauge theories, both the one-loop self-energy and
vertices must be resummed into
effective quantities [3]. Since these quantities are momentum dependent, the
 effective expansion is quite involved even at one-loop order. Nevertheless,
one expects that some of the features of two-loop calculations
studied here in a simpler context will also manifest
themselves in gauge theories.
\vfil
\newpage
{\bf Acknowledgements}\\
I gratefully acknowledge : Dr. R. D. Pisarski for suggesting the
 problem and for the many
 subsequent discussions and helpful comments; and Prof. M. Ro\v{c}ek for
 his valuable advice and support. I also thank Prof. J. C. Taylor
and the referee for pertinent questions and comments concerning
temperature dependent counterterms.
Finally, the author thanks C. Corian\`{o},
G. Estefan and R. Stewart  for inspiring and encouraging conversations.

This work was supported in part by NSF grant No. PHY 91-08054.
\vfil
\pagebreak

{\appendix
{\mysection{Appendix}
(1) The basic expression appearing in one-loop diagrams is \\
\be
\mu^{2 \epsilon} \mbox{Tr} \Delta_{k} (K) \equiv I_{0}(M) +
I_{\beta}^{\epsilon}(M)
\eq
with
\be
I_{0}(M) = \mu^{2 \epsilon} \meas {1 \over 2 E_{k}} =
 {M^2 \over (4\pi)^2} \left({4\pi \mu^{2} \over M^2} \right)^{\epsilon}\ \Gamma
(-1 + \epsilon)
\eq
\be
I_{\beta}^{\epsilon}(M) = \mu^{2 \epsilon} \meas {n_{k} \over E_{k}}
\eq
For $(M/T) \ll 1$, we have the expansion [13],
\be
I_{\beta}^{\epsilon =0}(M) = {T^2 \over 12}\left[1 -{3M \over \pi T} - {3
\over 4\pi^2}\left({M \over T}\right)^2 \ln \left({M \over T}\right)^2 - {3
\over \pi}\left({M \over T}\right)^2 \left({\gamma_E \over 2} - {\ln 4\pi
\over 2} - {1 \over 4}\right) \ldots \right] \nonumber \\
\eq

The expression  $\mu^{2 \epsilon} \mbox{Tr} \left[ \Delta(K) \right]^2 $ can be
obtained
by differentiating (A.1-3 ) with respect to $M^2$.

\noindent 2. To get (3.11-14).\\
Consider the $q$ integrals in (3.7) . For $\vec{p} = 0$ the only
 nontrivial angular integral in  $(D-1)$ dimensions
is for the angle $\theta$ between $\vec{k}$ and $\vec{q}$. Choose $\vec{k}$
to define the polar axis, and first do the trivial angular integrals for the
$q$-variables (see [9], for example, for the correct measure for the integrals
in $D$ dimensions). Then
one is left with the following integral $ (t= \cos \theta)$,

\be
\int_{0}^{\infty} {dq \over E_{q}} q^{1-2\epsilon} L(k,q)
\eq
where
\be
L(k,q) = \int_{-1}^{1} dt {1 \over(1-t^2)^{ \epsilon}} {\partial \over \partial
t}
\left ( \ln \left[ \omega^2 - (E_q + E_r - E_k)^2 \right] \left[ \omega^2 -
(E_q + E_r + E_k)^2 \right] \right)
\eq
The simplest way to proceed is to subtract the UV divergent
part of (A.5). As $ q \rightarrow \infty$ , ${\partial \over \partial t}
\left(  \right) \rightarrow 2k/q$. Subtracting and adding this term at the
appropriate place in (A.6), substituting everything back in (A.5), (3.7)
and then doing the obvious simplification  gives the result quoted in the text.

\noindent 3. Solving Eqn.(3.25).\\
Since $M^{2}_{3} \sim \vartheta (g^2)$, and $R(\omega^2) \sim \vartheta
(g^4)$, a
consistent way to solve (3.25) is by iterating the lowest order solution
$\omega_{0}^{2} = M_{3}^2$. The next iteration gives the result in the
text (3.26) while further iteration will give a correction
at $\vartheta(g^6)$.
Essentially then, the pole is determined by the self-energy on mass-shell. Now
consider a Taylor expansion of $R(\omega^2)$ about $\omega^2 =0$. Since $R$
has a logarithmic IR singularity as $\omega ,m \rightarrow 0$ (see text),
therefore

\be
R(\omega^2) = \sum_{n=0}^{\infty} {(\omega^2)^{n} \over n!} \ {\partial^n
R(\omega^2) \over \partial (\omega^2)^n } |_{\omega^2 =0} \sim g^4 \vartheta
\left( \sum_{n}
{\omega^{2n} \over (m^{2n}) n!} \right)
\eq
For $\omega$ soft ($\sim m$), the Taylor expansion is {\em not} an expansion in
 $g$ as  each term
is of order $g^4$. So one should expect
$ R(0)$ and $R(\omega^2)$ to differ by an amount $\vartheta(g^4)$ (see (3.23)).
The same argument explains why it would be  difficult to obtain the full order
$g^4$ contribution to the self-energy at soft non-zero external momentum
($\vec{p}$) by doing a Taylor expansion about $\vec{p} = 0$.

\noindent 4. To obtain (3.22).\\
To extract the leading ``$\ln g + $  constant'' contribution
from the integrals appearing in (3.14-15), the following procedure
is adopted : identify the terms which
will contribute the $\ln m$ singularities, isolate them and then set $m=0$
 in the
 regular terms to get pieces of the $\vartheta(g^4)$ contribution. Next,
 simplify the potentially singular terms and keep repeating the
 above procedure until
all the ``log + constant '' pieces have been explicitly obtained.
Consider for example $F_{2}(\omega^2 =M_{3}^2)$. The term within braces in
(3.14) is
simplified (for $\omega$ on-shell) and written  (replacing $M_{3}$
 with $m$ in the expressions  $F_{2}$  and $H$ ignores a correction of
 order $g^5$) as follows

\bea
& & q \ln \left[ \left({q +k \over q - k} \right) \left( {m^2 + q(q+k) +
E_{q}E_{q+k} \over T^2 } \right)\right] + \\
& & \; -q \ln \left| {m^2 + q(q-k) + E_{q}E_{k-q} \over T^2} \right| + \\
& & \: -4k
\eea
As discussed in the text, there are no IR singulrities in the region
$q \geq k$ , so one may set $m=0$ in the above expressions in that range of
integration. For the $q \leq k$ sector, (A.9) and (A.10) both contribute
`logs'
while (A.8) gives a finite piece in the massless limit. For the piece in
(A.10),the $q$ integral is easily done explicity, then the  $\ln m$ piece
isolated
 and the finite terms determined. For (A.9), multiply the argument of the
logarithm by  $(m^2 +q(q-k) - E_{q}E_{k-q})$ in the numerator and denominator,
 simplify, isolate the
`log' piece and set $m=0$ in the rest to get the constants. Collecting  all the
terms gives
\be
F_{2}(M_{3}^{2}) = {g^4 \over 6}{T^2 \over 64 \pi^2} \left[ {1 \over 2} \ln
({m \over T})^2 -\ln 2 - {6 \over \pi^2} \int_{0}^{\infty} dx {x \ln x \over
e^x -1} \right]  + \vartheta(g^5 \ln g) \\
\eq
The integral in the final answer (A.11) is a pure number. It  may be computed
numerically if required. The concise result is given in the text (3.22).

For a different example, consider

\be
H(M_{3}^{2}) = 2 { g^4 \over 64 \pi^4} \int_{0}^{\infty} dk{k n_{k}
\over E_{k}} \int_{0}^{k} dq {q n_{q} \over E_{q}} \ln \left|{k+q \over k-q}
 \right| ,
\eq
where the symmetry of the integrand under $k \leftrightarrow q$ interchange
has been exploited to restrict the range of one of the integrals. As the IR
singularity is now due to the extra BE factor rather than the explicit
logarithm, the `log + constant' pieces can be obtained
by using an arbitrary soft
cutoff $\Lambda $ (I thank R. D. Pisarski for suggesting this technique)
to divide the region of integration for the
 $k$ variable. For $k \leq \Lambda$, the appropriate approximations can be
 made ( e.g. $n_{k} = 1/E_{k}$) to simplify the
integrations . In the limit $a \rightarrow 0$, one gets for $k \leq \Lambda$
in (A.12),

\be
T^2 \left[ {- \pi^2 \over 4} \ln ({m \over T}) + { \pi^2 \over 4}
\ln ({\Lambda \over T}) + \int_{0}^{1} dt \ln \left| {1+t \over 1-t} \right|
 {\ln t \over (t - t^3)} \right] + \vartheta(m) + \vartheta(\Lambda)
\eq
For the region $k \geq \Lambda$, one can put the mass to zero because there
are no $\ln m$ singularities. Next isolate the $\ln \Lambda $ factor (for
$ \Lambda \rightarrow 0$) by doing a subtraction of the leading IR part of one
of the BE factors, to get finally for $k \geq \Lambda$ in (A.12),
\be
T^2 \left[-{\pi^2 \over 4} \ln ({\Lambda \over T})+ \int_{0}^{1} dt
\int_{0}^{\infty} dx
{x \over e^x -1}( {1 \over e^{xt} -1} - {1 \over xt})
\ln \left|{1+t \over 1-t} \right|\right] + \vartheta(m) + \vartheta(\Lambda)
\eq
In the limit $\Lambda \rightarrow 0$ , the sum of (A.13) and (A.14) gives
the final answer for (A.12), with neglected terms of order $g^5$.
The cancellation of the $\ln \Lambda$ terms in the sum removes the ambiguity
coming from the
cutoff.

Similar considerations as above give

\be
F_{2}(0) = - {g^4 \over 6}{T^2 \over 64 \pi^2}{\pi \over \sqrt{3}}
+ \vartheta(g^5) ,
\eq

\bea
H(0) &=& -{g^4 \over 6}{T^2 \over 64 \pi^2} \ln({m \over T})^2
\nonumber \\ \nonumber\\
& & \ + {g^4 T^2 \over 32 \pi^4} \int_{0}^{1} dt {\ln t \over  (t-t^3)}
\ln \left({1+t+t^2 \over 1-t+t^2} \right) \nonumber \\
\\
& & \ + {g^4 T^2 \over 32 \pi^4} \int_{0}^{1} dt \int_{0}^{\infty} dx
{x \over e^x -1}( {1 \over e^{xt} -1} - {1 \over xt})
\ln \left({1+t+t^2 \over 1-t+t^2} \right) \nonumber \\
\nonumber \\
& & \ + {g^4 T^2 \over 32 \pi^4} \int_{0}^{1} dt \int_{0}^{\infty} dx
{t x^3 \over (x^2 + 1)(x^2 t^2 + 1)}
\ln \left[{4 x^2 + {3 \over (1+t+t^2)} \over 4 x^2 + {3  \over (1-t+t^2)}}
 \right] \nonumber \\
 & & \; + \vartheta(g^5 ). \nonumber
\eea

Again, if necessary, the constant integrals appearing  above can be done
numerically to give the result in (3.22).

\vfil
\pagebreak

\noindent{\Large{\bf References}}

\begin{itemize}
  \item [1] A. Linde , {\it Rep. Prog. Phys.} {\bf 42} (1979) 389;\\
            D. J. Gross, R. D. Pisarski and L.G. Yaffe,
            {\it Rev. Mod. Phys.} {\bf 53} (1981) 43.

  \item [2] S. Weinberg, {\it Phys. Rev. D} {\bf 9} (1974) 3357.

  \item [3] R. D. Pisarski,  {\it Phys. Rev.  Letts.} {\bf 63} (1989) 1129;
            E. Braaten and R. D. Pisarski,  {\it Nucl. Phys. B} {\bf 337}
            (1990) 569; R. D. Pisarski, {\it Resummation and the Gluon
   Damping Rate in Hot QCD}, talk presented at Quark Matter'90; Menton,
   France,    May 1990.

  \item [4] See for example, E. Braaten and T.C. Yuan,
           {\it Phys. Rev.  Letts.} {\bf 66} (1991) 2183,
            and the references therein.

  \item [5] T. Altherr, {\it Phys. Lett. B} {\bf 238} (1990) 360.

  \item [6] N. Banerjee and S. Mallik, {\it Phys. Rev. D} {\bf 43} (1991) 3368.

  \item [7] S. P. Chia, {\it Int. J. Mod. Phys. A} {\bf 2} (1987) 713;\\
         P. Fendley, {\it Phys. Lett. B} {\bf 196} (1987) 175.

  \item [8] M. B. Kislinger and P. D. Morley, {\it Phys. Rev. D} {\bf 13}
 (1976) 2771;\\
   H. Matsumoto, I. Ojima and H. Umezawa, {\it Ann. Phys. (NY)}
 {\bf 152} (1984) 348.

  \item [9] B. De Wit and J. Smith, {\it Field Theory in Particle Physics}
(North-Holland, Amsterdam 1986).

  \item [10] A. A. Abrikosov, L. P. Gorkov and I. E. Dzyaloshonski,
  {\it Methods
  of Quantum Field Theory in Statistical Physics} ( Dover, New York 1975).

  \item [11] J. Kapusta ,{\it Finite Temperature Field Theory} (Cambrige
 University Press, 1985).

  \item [12] R. D. Pisarski , {\it Nuc. Phy. B} {\bf 309} (1988) 476.

  \item [13] L. Dolan and R. Jackiw, {\it Phys. Rev. D} {\bf 9} (1974) 3320.

  \item [14] H. A. Weldon, {\it Phys. Rev. D} {\bf 28} (1983) 2007.

  \item [15] E. Mendels, {\it Il Nuo. Cim.} {\bf 45} (1978) 87.

  \item [16] It is well known that no temperature dependent counterterms are
required in bare perturbation theory [8]. The same is true for the net
lagrangian (even with a possible nonzero $T=0$ mass) in the resummed expansion
if a mass-independent renormalisation scheme is used. The latter fact is
transparent in the approach used in [6].

\vfil
\pagebreak

\noindent{\underline{\Large Figure Captions}}
\bigskip

\underline{Fig.1}:\\ Fig.1a is the one-loop self-energy diagram,
also called the ``bubble''. Fig.1b shows the ultraviolet mass counterterm
while Fig.1c is the finite 2-point interaction (``blob'') counterterm
induced by the resummation.

\underline{Fig.2}: \\Fig.2 is the
``daisy'' diagram with $N \geq 1$ attachment of bubbles.

\underline{Fig.3}:\\  The the set of $ N =1$ daisy-like
diagrams; Fig.3a is the $N=1$ daisy while Fig.3b has an insertion of the
finite 2-point interaction (blob) into the one-loop self-energy (bubble)
diagram.

\underline{Fig.4}:\\
The set of $N=2$ daisy-like diagrams.

\underline{Fig.5}: \\
A general daisy-like diagram with $N \geq 1$ bubbles + blobs attached.

\underline{Fig.6}:\\ Fig.6a is the one-loop correction
to the 4-point vertex, the diagrams in the crossed channels are not shown.
Fig.6b is the UV vertex counterterm.

\underline{Fig.7}: \\ Renormalisation counterterms for second
order calculations. Fig.7a is the counterterm for Fig.3b. The mass and vertex
counterterms are  Figs.7b and (7c) respectively.

\underline{Fig.8}: \\Overlapping two-loop self-energy diagram.

\underline{Fig.9}:\\
Diagrams which are individually of order $g^4$ but sum to higher order.

\underline{Fig.10}: \\
A general ``cactus'' diagram. $N \geq 0$ bubbles  + blobs are attached
to the middle loop while $M \geq 0$ bubbles + blobs are attached to the
bottom loop.

\underline{Fig.11}: \\
More complicated self-energy diagrams which are individually of order $g^4$
but sum to higher order.

\vfil

\pagebreak

\end{itemize}

\end {document}